\pdfoutput=1

\documentclass[11pt]{article}

\usepackage{EMNLP2022}
\usepackage{times}
\usepackage{latexsym}
\usepackage{graphicx}
\usepackage[T1]{fontenc}

\usepackage[utf8]{inputenc}

\usepackage{microtype}

\usepackage{inconsolata}

\title{Detecting Chinese Fake News on Twitter during the COVID-19 Pandemic}

\author{Yongjun Zhang  \\
  Stony Brook University \\
  \texttt{yongjun.zhang@stonybrook.edu} \\ \AND
  Sijia Liu  \\
  Amazon.com, Inc \\
  \texttt{sijial@amazon.com} \\ \AND
  Yi Wang \\
  Stony Brook University \\
  \texttt{yiw.wang@stonybrook.edu} \\ \AND
  Xinguang Fan \\
  Peking University \\
  \texttt{xfan19@pku.edu.cn} 
  }

\begin{document}
\maketitle
\begin{abstract}
The outbreak of COVID-19 has led to a global surge of Sinophobia partly because of the spread of misinformation, disinformation, and fake news on China. In this paper, we report on the creation of a novel classifier that detects whether Chinese language social media posts from Twitter are related to fake news about China. The classifier achieves an F1 score of 0.64 and an accuracy rate of 93\%. We provide the final model and a new training dataset with 18,425 tweets for researchers to study fake news in the Chinese language during the COVID-19 pandemic. We also introduce a new dataset generated by our classifier that tracks the dynamics of fake news in the Chinese language during the early pandemic.
\end{abstract}

\section{Introduction}

Alongside the COVID-19 pandemic, the spread of misinformation, disinformation, and fake news via social media platforms such as Twitter, Facebook, and Tiktok has led to the infodemic concerned by policymakers, journalists, academics, and the public \citep{buchanan2020managing,germani2021anti,simon2021autopsy,zarocostas2020fight}. Scholars have developed a series of computational tools to detect and monitor COVID-19 fake news on social media platforms, but prior studies focus overweeningly on the dominant languages in Western societies such as English and German \citep{bang_model_2021,mattern_fang-covid_2021,vijjali_two_2020}. 

We focus on fake news detection in the Chinese language context on a western social media platform in this paper, given that China has become the epicenter of this infodemic due to the conspiracy on the origin of COVID-19 from a Wuhan lab. During the early pandemic, the misinformation and fake news on COVID-19 was first spread within the Chinese community and then diffused to other communities. However, we lack critical data to quantify and understand the scale and scope of the diffusion of fake news related to China during the early COVID-19 pandemic in its own community.

In this paper, we adopt a subjective opinion-based approach to detect fake news on Twitter in 2019-2020 \citep{zhang_fake_2020}. Unlike the fact-check approach relying on third-parties (e.g., news media, government organizations, experts), we mimic the real world scenario about how fake news recipients perceive these messages and focus on social media users' subjective opinion of whether or not these messages are fake news. 

Our contribution is threefold. First, we provide a unique annotated training dataset with over 18,000 tweets. To our best knowledge, this is the first training dataset in the community specifically related to fake news on China during the pandemic. Second, we fine-tune a series of transformer models and train our own classifiers that can be used to classify whether a Chinese tweet is fake news or not. Third, we also provide a dataset with over 25 million Chinese tweets obtained via Twitter academic API using keywords mentioning China, Chinese, CCP, etc. We provide an overview of the dynamics of fake news on Twitter in the Chinese language since the outbreak of COVID-19. 

\section{Related Work}

In recent years, both industrial and academic research scientists have devoted great attention to fake news detection due to the growing concerns regarding the adversarial consequences of fake news. As a result, multiple large-scale annotated fake news datasets have been created for algorithm training, but they are primarily in the English language \citep{mattern_fang-covid_2021,shu_fake_2017,shu_fakenewsnet_2019,shu_mining_2020}. This allows scholars to train state-of-the-art deep learning models to detect and monitor fake news, and these methods account for both texual and contentual information (e.g., social media data) into fake news detection \citep{shu_studying_2019,zhou_fake_2020,zhou_network-based_2019}. Next, we provide a brief overview of prior fake news research related to the concept of fake news, approaches of fake news detection, and datasets of fake news detection.

\subsection{Defining Fake News}
An accurate and operational definition of fake news is a big challenge for fake news research. Existing studies adopt various concepts and terms, such as clickbait \citep{chen_misleading_2015}, misinformation \citep{lazer2018science}, disinformation \citep{mattern_fang-covid_2021}, and false information \citep{shu_fake_2017}. \citet{egelhofer_fake_2019} offer a two-dimensional framework to define fake news: fake news genre (describing the deliberate creation of pseudojournalistic disinformation) and fake news label (the instrumentalization of the term to delegitimize news media). \citet{zhang_fakedetector_2020} survey the fake news literature and provide two different definitions. For the broad definition, fake news is simply false information, while the narrow one restricts fake news within the intentionally false news published by news outlets. The underlying consensus behind current fake news detection techniques takes the broad approach that fakes news is at least false information.

\subsection{Approaches to Studying Fake News}
Methods of fake news detection can be summarized into three main approaches. \textit{Fact-checking} approach relies on the text of questionable claims and trains models on labeled claims from online fact-checking resources, such as \textit{classify.news} and \textit{FactCheck.org} \citep{graves_understanding_2018,pathak_breaking_2019,popat_declare_2018,rashkin_truth_2017}. This approach does not use explicit features and often fails to offer interpretable explanations. \textit{Style-classification} approach focuses on the textual (context texts, images, videos, etc.) and contextual features (user profiles, user posts and responses, etc.) of news and uses machine learning or deep learning based models to detect fake news \citep{przybyla_capturing_2020,zhang_fakedetector_2020}. However, two issues arise from this approach: dependence on context and news contents. Dependence on context indicates that fake news style varies by time, languages, and regions, while heavy dependence on news contents may inspire countermeasures by fake news makers. \textit{Network-based} approach relies on news cascade and spreading features of fake news \citep{liu_fned_2020,zhang_fake_2018,zhou_network-based_2019,zhang_fake_2020,shu_studying_2019}. This approach uses direct or indirect information on news propagation to train traditional machine learning or deep learning models and, therefore, mainly applies to the task of detecting fake news which has been spread over a period of time. Other scholars often categorize different approaches into knowledge-based and propagation-based fake news detection as well. The former focuses more on determining the truthfulness of news time while the latter emphasizes the credibility of news sources.

\subsection{Fake News Benchmark Datasets}
A few benchmark general fake news datasets have been released publicly in recent years. For instance, several large-scale fake news related datasets have been used by scholars to build fake news detection algorithms, including \textit{BuzzFeedNews}, \textit{LIAR}, \textit{BS Detector}, \textit{CREDBANK}, \textit{BuzzFace}, \textit{FacebookHoax},\textit{FakeNewsNet} (see \citealp{shu_fakenewsnet_2019} for a detailed review). These datasets focus on different dimensions of fake news, such as linguistic, visual , social , and spatiotemporal features. 

To combat the COVID-19 infodemic, scholars have developed a series of COVID-19 specific fake news datasets, including \textit{Covid19 FN} \citep{patwa2021fighting}, \textit{FakeCoovid} \citep{shahi2020fakecovid}, \textit{ReCOVery} \citep{zhou2020recovery}, CoAID \citep{cui2020coaid}, and \textit{FANG-COVID} \citep{mattern_fang-covid_2021}. Of course, these datasets are overwhelmingly in English \citep{hangloo_fake_2021}, except for \textit{FANG-COVID} and \textit{FakeCovid}. Given that content style varies across languages and societies, non-English data availability is crucial for the further development and application of fake news detection techniques.

\subsection{Fake News Detection }
Scholars have developed a variety of machine-learning or deep learning models to detect fake news (see \citealp{patwa2021fighting} for a more comprehensive review). Here we briefly describe some models trained on the above datasets we have mentioned. For instance, \citet{glazkova2021g2tmn} fine-tuned Covid-Twitter-BERT model \citep{muller2020covid} and achieved an F1 score of 0.987. \citet{shahi2020fakecovid} obtained an F1-score of 0.78 based on English articles from FakeCovid with a pretrained BERT model. \citet{zhou2020recovery} trained a neural network that combines textual and visual information and achieved an F1-score of 0.672 for the detection of fake news using ReCOVery fake news data. \citet{mattern_fang-covid_2021} trained a deep learning model combining BERT with social context information based on \textit{FANG-Covid} dataset with 41,242 German news articles and associated twitter information and achieved an accuracy of 0.981 and an F1 score of 0.966. But other models trained on general fake news datasets are not quite effective, which warrants further research on specific topics.

\section{Introducing the Datasets}

Next, we introduce two datasets used in this paper: CNTweets data and CNFakeTweets data. The former contains all Chinese tweets mentioning China, Chinese, Chinese Communist Party, Chinese government, etc. The latter is the training dataset we compiled for fake news detection on Twitter.

\subsection{CNTweets Data}
To evaluate how fake news in the Chinese language was diffused across communities, we rely on the CNTweets, a dataset containing all Chinese tweets mentioning China-related keywords, such as China, Chinese, Chinese Communist Party, etc. This dataset contains over 25 million tweets by 1.32 million users scraped from the Twitter historical database using academic API. CNTweets data has 16.84 million tweets mentioning China and 8.39 million mentioning CCP.

\subsection{CNFakeTweets Data}

To build our training dataset \textit{CNFakeTweets}, we started with those Twitter users who are most/least likely to post fake news related to China and their followers or following accounts, and these accounts typically either support or criticize the Chinese government or China. We obtained all their tweets posted in the past 2 years. We also used pro- and anti-China hashtags and keywords to obtain potential tweets. We then used a stratified sampling strategy to select 12,000 tweets. To add more potential neutral tweets in our training dataset, we also randomly selected 3,000 tweets from the CNTweets data. These 15,000 tweets were annotated by at least two research assistants.

We then use the annotated training dataset to train our algorithm to classify fake news in our CNTweets database. We take a semi-supervised machine learning approach to construct our final training dataset. After fist-stage of classification, we annotate another random sample of 5,000 tweets classified by our algorithm as fake news. We then feed these tweets into our training dataset for our final classifier building. Thus, our final training dataset contains 18,435 tweets after data cleaning with 2,599 labeled fake news.

\section{Data Annotation}
We adopt a subjective opinion-based approach to annotate fake news in this study. We broadly define fake news as false claims that mimic news information and written intentionally to mislead readers \citep{shu_fake_2017,zhang_fake_2020}. We ask research assistants to read the content of a tweet solely based on its text and then evaluate whether it is fake news related or not based on their subjective reading and personal experiences. All tweets are evaluated at least by two annotators and we treat it a tweet as fake news as long as one annotator labels it as fake news.

\section{Model Training and Performance}

\subsection{Transfer learning and fine tuning}

We take a transfer learning approach to build our models to identify fake news in the Chinese language on Twitter during the early pandemic. Since previous studies consistently show that transformer models have outperformed other language models, we fine-tune a series of BERT models trained on Chinese corpora to categorize a tweet into fake news or not. We compare the results based on Chinese-BERT-WWM-ext, Chinese-RoBerta-WWM-ext, and Chinese-MacBert-large.

\subsection{Model performance}

Table~\ref{model-performance} reports our model metrics. Our pioneer work on detecting fake news on Twitter in the Chinese language offers scholars some baselines for comparison. Note that we only account for textual information in a tweet. 

\begin{table}[htb]
\centering
\begin{tabular}{lcccc}
\hline
Model Name & Accuracy&F1 score\\
\hline
Chinese-bert-wwm-ext&0.926&0.656\\
Chinese-roberta-wwm-ext&0.932&0.663\\
Chinese-macbert-large&0.929&0.648\\
\hline
\end{tabular}
\caption{\label{model-performance}
Transformer Model Performance
}
\end{table}

\section{The Overview of Chinese Fake News on Twitter}
Next, we present the overview of fake news in the Chinese language on Twitter during the early pandemic from 2019 to 2021. We use our best model to classify CNTweets data. Among over 25 million tweets, we find 0.97 million tweets classified as fake news. Roughly 4\% of tweets posted in Chinese are classified as fake news.

\subsection{Time Series of Fake News in Twitter}
\begin{figure}
    \centering
    \includegraphics[width=.5\textwidth]{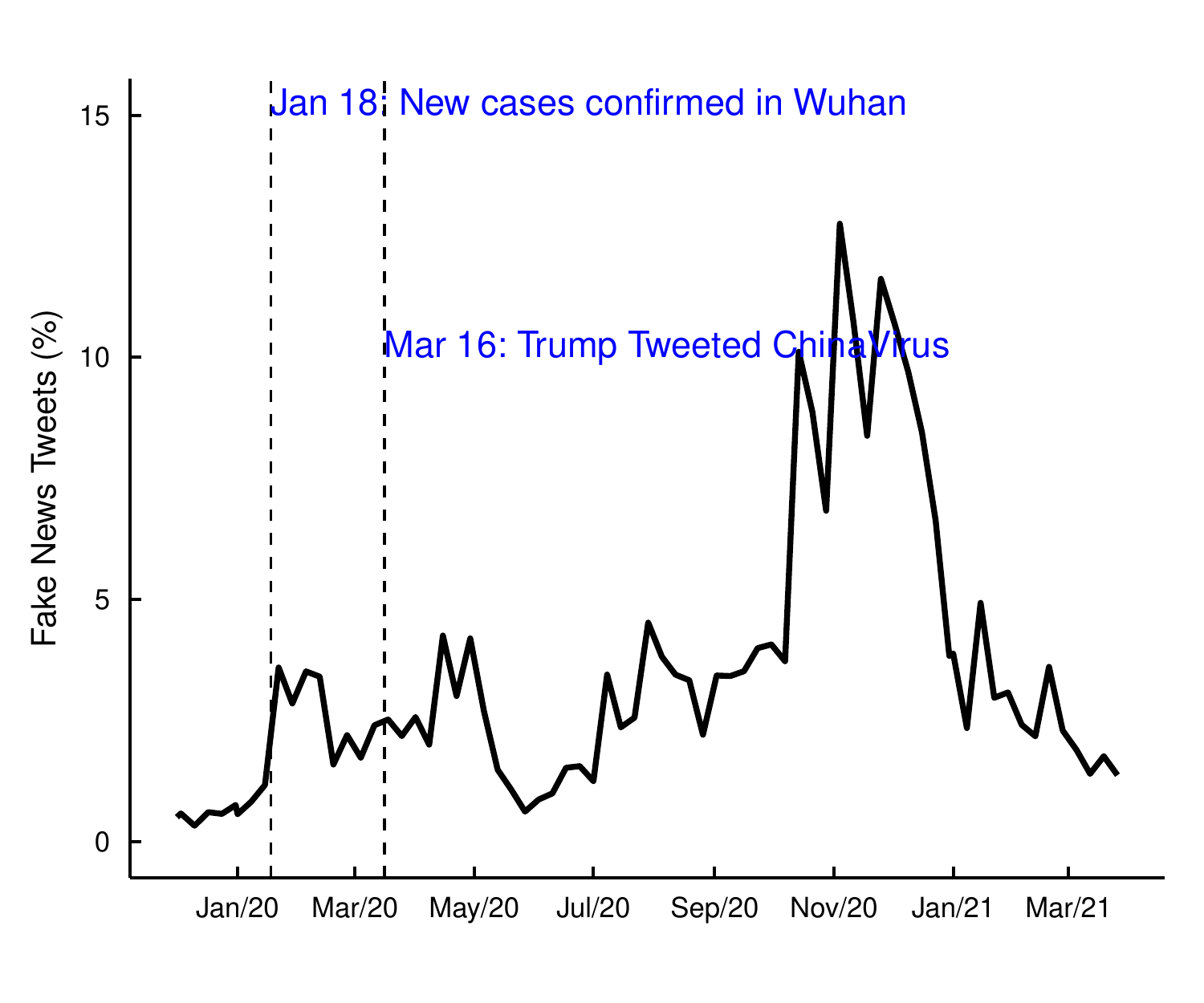}
    \caption{Weekly Trend of Tweets Labeled as Fake News in CNTweets Database}
    \label{fig:fktrend}
\end{figure}

Fig.~\ref{fig:fktrend} shows the weekly trend of fake news in our CNTweets database. Overall, fake news persisted during the early pandemic. The early two spikes around February and May in 2020 were due to the conspiracy theory on the origin of COVID, and fake news on the Wuhan Lab was shared and retweeted by many Twitter users. The false information peaked around late 2020 and early 2021. Roughly 13\% of tweets in our CNTweets database were classified as fake news. This is because misinformation on U.S. presidential election was spread on Twitterverse, for instance, the Hunter Biden laptop controversy.

\subsection{Content of Fake News on Twitter}
\begin{figure}
    \centering
    \includegraphics[width=.5\textwidth]{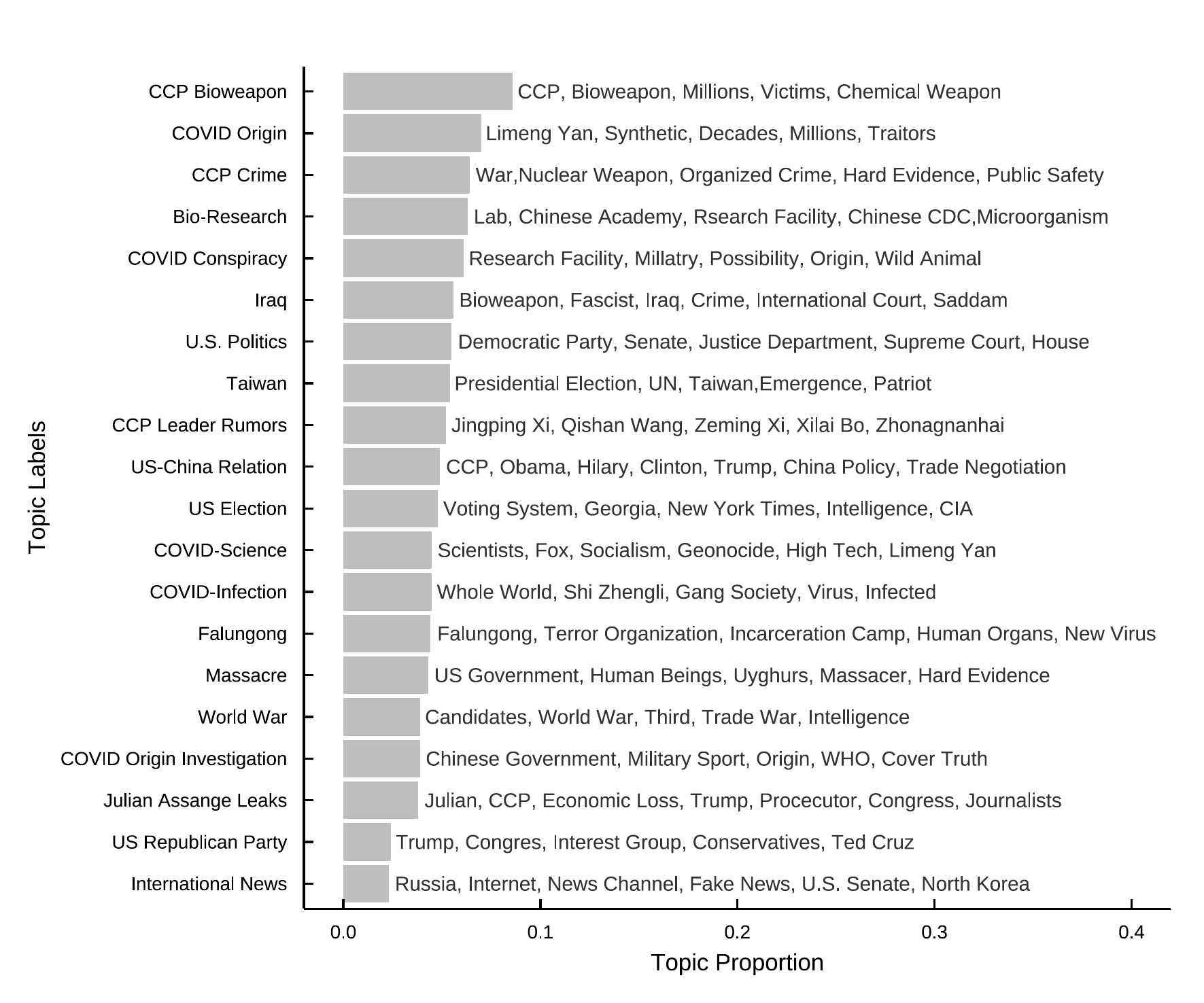}
    \caption{Structural Topic Model Results (K=20)}
    \label{fig:fktopic}
\end{figure}

To further explore the contents of these fake news on Twitter, we run structural topic modeling to extract the major themes \cite{roberts2019stm}. Fig.~\ref{fig:fktopic} shows the explored 20 topics in our database. The most prevalent topic in the extracted fake news pertains to COVID-19 conspiracy theories (e.g., made in Wuhan Lab and Limeng Yan's interview by Fox News) during the early pandemic. In addition, fake news posted in Chinese associated with U.S. presidential election warrants future scholarly attention. False information related to both Republican and Democratic candidates in 2020 accounted for a large proportion in our fake news database. 

\section{Conclusion}
Computer and social scientists have devoted great efforts to detect misinformation and fake news on social media platforms in recent years. Our paper provides the fist attempt, to our best knowledge, to systematically detect fake news in the Chinese language on a social media platform, i.e., Twitter. We annotated a unique training dataset with over 18,000 tweets used to build a transformer-based classifier. We then employed this classifier to a large-scale dataset during the early pandemic containing tweets mentioning any keywords related to China, Chinese, and Chinese Communist Party in the Chinese language. Our data visualization shows a rise and fall of fake news on Twitter within the Chinese language community that needs further studies.

This paper adopts a subjective opinion-based approach to evaluate the credibility of a tweet, and our approach captures the subjective assessment of whether a tweet is a news-like false claim. Future research can further conduct fact-checking to verify the reliability of these subjective rating. In addition, this study does not incorporate Twitter-user or post-level auxiliary information, such as user profile, followers, following, and number of likes, into modeling. Future research can further improve our classifier by retraining BERT models with extra social media information.

\section*{Limitations}
Our research has several limitations as well. First, our classifier only considered textual information in a tweet. As tweets posted by users are multimodal, researchers can take images, emojis, and videos into account in the future research. Second, previous research has shown that Twitter user information has substantially improved the accuracy of fake news classification. Although our work does not incorporate authors' information, we have collected users' timeline information and their networks and we will address this issue in the future work. Third, our classifier might not be extended to other periods since it focused on the early COVID pandemic. One approach to extend this line of work is to add more training examples in our CNFakeTweets dataset. Lastly, we took an subjective approach to classify fake news, and one future direction would be to combine subjective approach with fact-check. It would be also important to compare the inconsistency between different annotators with intercultural backgrounds.

\section*{Ethics Statement}
We used Twitter academic API to collect our CNTweets and CNFakeTweets datasets. Due to Twitter's term of service, we cannot share the original tweets we have collected, but we can share Tweet IDs and our classifiers. In addition to data ethics, our work also raises other ethical concerns. First, the training dataset labeled by research assistants have a lot of misinformation and racial slurs, which could cause some mental health issues. Scholars should prepare annotators with enough background and resources when they annotate the training dataset. Second, our work only reveals the overall pattern in our collected database and these results do not reflect our personal opinions.


\bibliographystyle{acl_natbib}

\end{document}